# Results of numeric integrations of orbits of (5026) Martes and 2005 WW113 asteroid pair

Rosaev A.

Research and Educational Center "Nonlinear Dynamics", Yaroslavl State University

The orbital dynamic of very young asteroid pair **(5026) Martes and 2005 WW113** is studied by numerical integrations. We discover strong resonant perturbations of the larger member of pair **(5026) Martes** by the mean motion resonance 3:11 with Earth. The unbounded secondary (**2005 WW113**) moved far from the resonance and is not perturbed.

Introductions

The close and young asteroid pairs are intensively studying recently (Pravec et al., 2019) [1]. The study of such pair is more easy then old and dispersed asteroid families. The present paper is devoted to investigation of one of such pair (5026) Martes and 2005 WW113. Previously we note that this pair orbited close to 3:11 mean motion resonance with Earth [2]. In this paper we present results of our numerical integration of the pair and more detailed study of resonant perturbations.

Some examples of the resonance perturbations of young asteroid families and pairs are known, first of al, it is Datura family in 9:16 resonance with Mars (Nesvorny, et al, 2006) [3]. Pravec et al (2019) [1] note that pair (49791) 1999 XF31 and (436459) 2011 CL97 chaotic orbits may be explained by 15:8 mean motion resonance with Mars. Duddy et al (2012) [4] pointed that pair (7343) Ockeghem and (154634) 2003 XX38 is in 2-1J-1M three body resonance.

As it was noted in paper Broz and Vokrouhlicky (Broz M., Vokrouhlicky D.,2008) [5], in resonance we have monotonic changes in eccentricity instead semimajor axis drift. As a result, resonances can occur significant effect on evolution some close asteroid pairs.

Throughout the article we have used standard notations for orbital elements $a$ – semimajor axis in a.u., e – eccentricity, $i$ – inclination, n – mean motion, $\Omega$ – longitude of ascending node, $\omega$ – perihelion argument, $\varpi$ – longitude of perihelion (the angular elements are in degrees).

## (5026) Martes and 2005 WW113 pair main facts

This pair studied by Pravec et al (2019) [1], who found that this is a very young pair. Their backward integrations of heliocentric orbits suggest that these two asteroids separated about 18 kyr ago.

The proper orbital elements of the pair get in the AstDys site [6] are given in the Table 1.

Table 1. Proper elements of pair 5026 Martes – 2005 WW113. Data of elements 17.11.2023

| Asteroid | Synthetic $a$ | Analytic $a$ | e | LCE | g | s |
|---|---|---|---|---|---|---|
| 5026 Martes | 2.37752 | 2.37739 | 0.208072 | 40.57 | 38.7909 | -46.0335 |
| 2005 WW113 | 2.37752 | 2.37698 | 0.207076 | 24.92 | 38.759 | -45.9692 |

Pair 5026 Martes – 2005 WW113 is an example, when analytic proper elements have an advantage: synthetic proper semimajor axis of both asteroid in the pair are equal, when in reality they are different in the observed time interval 80 kyr: it is rightly reflected in analytic proper elements (table 1, Fig.1). Of course, in the long time interval, after the number of chaotic jumps, semimajor axis of 5026 Martes and 2005 WW113 may become very similar.

The osculating orbital elements of the pair are given in the Table 2 and 3 where orbit of 5026 Matres is based on 1758 and 3199 observations respectively. The uncertainties of orbits of 5026 Martes and 2005 WW113 were given in Table 4. The precision of the orbit determination of 5026 Martes is about five times better than 2005 WW113.

Table 2. Osculating orbital elements at **Epoch 06.07.1998 (JD 2451000.5 TDB). (**Data of elements 2019-Jul-21**)**

| Asteroid | $\omega$, deg | $\Omega$, deg | $i$, deg | $e$ | $a$, AU | $M$, deg |
|---|---|---|---|---|---|---|
| 5026 Martes | 16.78112050 | 305.0502029 | 4.3012146 | 0.24360765 | 2.37684843228 | 354.083998 |
| 2005 WW113 | 16.42208798 | 305.1261504 | 4.2951311 | 0.242305464 | 2.37645499298 | 73.7147646 |

Table 3. Osculating orbital elements at **Epoch 06.07.1998 (JD 2451000.5 TDB). (**Data of elements 2023-Oct-01**)**

| Asteroid | ω, deg | Ω, deg | *i*, deg | *e* | *a*, AU | *M*, deg |
|---|---|---|---|---|---|---|
| 5026 Martes | 16.78109387 | 305.0502492 | 4.3012144 | 0.243607648 | 2.37684843358 | 354.083989 |
| 2005 WW113 | 16.42219448 | 305.1260679 | 4.2951305 | 0.242305450 | 2.37645495301 | 7.37147118 |

Table 4: Uncertainties of orbital elements (1σ interval)

| Asteroid | | δω,deg | δΩ,deg | δ*i*, deg | δ*e* | δ*a*, AU | δM, deg |
|---|---|---|---|---|---|---|---|
| 5026 | Martes | 5.206e-05 | 5.148e-05 | 4.608e-06 | 4.286e-08 | 6.744e-09 | 7.861e-06 |
| | 2005WW113 | 0.0002654 | 0.0002647 | 1.024e-05 | 1.523e-07 | 3.31e-08 | 3.946e-05 |

**Method and problem setting**

To study the dynamical evolution of these close asteroid pairs, the equations of the motion of the systems were numerically integrated 50 kyrs into the past, using the N-body integrator Mercury (Chambers, 1999) [7] and the Everhart integration method (Everhart, 1985) [8]. On base of [1] age estimation, we expect that this time interval is sufficient.

We made two series of integration of nominal orbit. In the first we use only the large planets perturbations. In the second we add the effect of Ceres, Vesta, Juno and Pallas. Then, to study the dependence of our results on initial conditions, we integrate a few clones of orbits. In the final series of our numeric integration we try estimate non-gravitational Yarkovsky effect. In addition to the calculation of the distance during encounter, we perform calculation relative velocity along integration by method of our previous paper **[9]**.

To the nominal resonance position calculation, we use values of semimajor axis of planets, averaged over time of integration: 1.52368 AU for the Mars, 5.20259 AU for the Jupiter, 1.000001 AU for the Earth.

To study interaction of the considered pair with the resonance and to determine position of the resonance center (chaotic zone center) we apply the integration of orbits of the asteroid with the significant values of Yarkovsky effect ($A2=1*10^{-13}$) and the different planetary perturbations.

**Results of nominal orbits integration**

There are two close encounters in pair on considered time interval. By using initial data in table 2 we obtain the following values. First one at epoch -2.68 kyr with minimal distance about 0.00028 AU, second at epoch **-18.58** kyr with minimal distance about 0.00053 AU (~ 80 000 km).

Hill sphere radius of Martes is about 760 km. Obviously, the obtained value is very large. However, orbital elements show clear convergence about 15 kyr ago (Fig.1) so, epoch 18.58 kyr is more preferred. It is confirmed by relative velocity calculations.

The synodic period between close encounters in pair is about 16 kyr and recent approaches take place about 2.6 and 18.58 kyr (Fig.2). When integrate orbits in future by interval 8 Myr, we observe fast divergence of angular orbital elements and according relative velocity increasing. Moreover, due different periods of node and perihelion longitude variations (different proper frequencies), so small difference in orbital elements never was repeated in the past. By this reason, epoch close to 18 kyr is most probable moment of forming of the pair. However, formally it is necessary to consider two other possibilities close to 2.6 and 34 kyr.

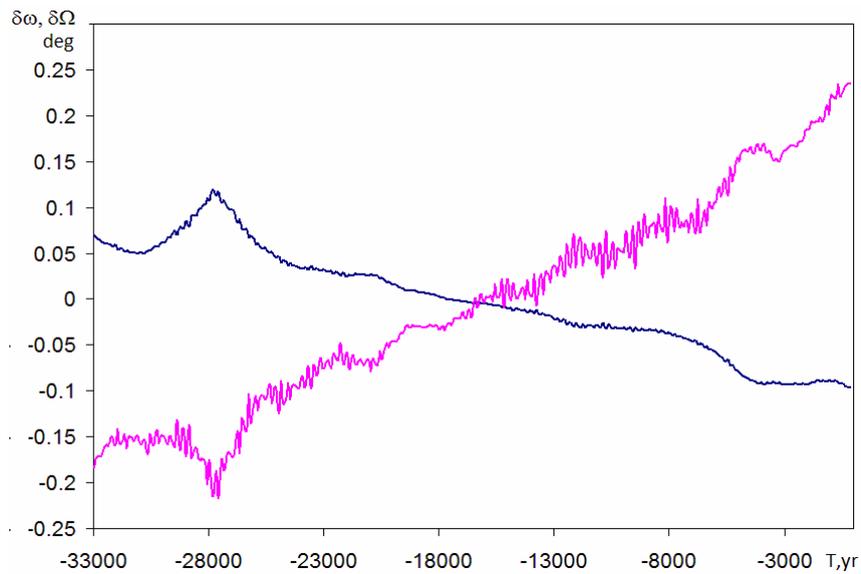

Fig.1. The differences of node and perihelion longitudes in pair **5026 Martes** – 2005 WW113

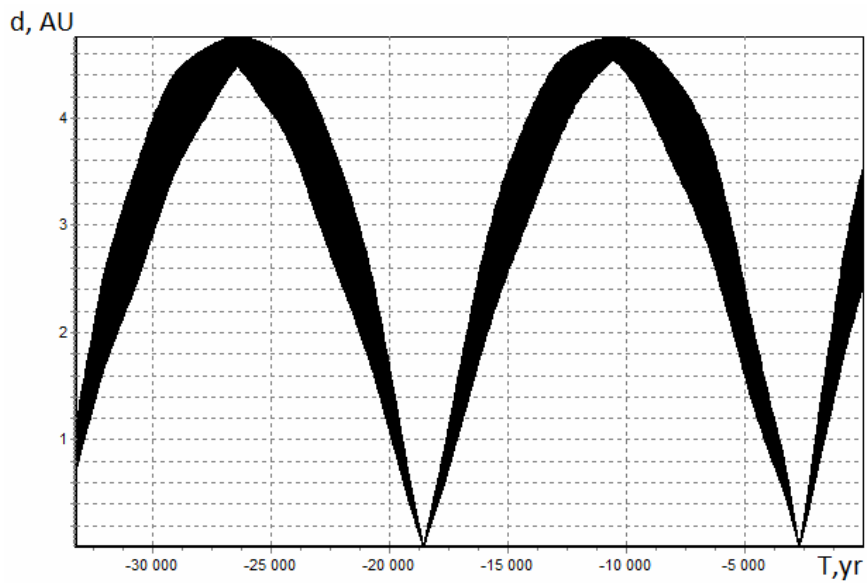

Fig.2. Synodic period in pair **5026 Martes** – 2005 WW113

**Resonance perturbations of pair (5026) Martes and 2005 WW113**

The main asteroid of pair 5026 Martes shows the strong resonance semimajor axis perturbations which are absent in the 2005 WW113 motion (Fig.3). This resonance may play the important role in the pair origin. By this reason, we consider it in the more details.

Resonance has significant width about 0.0002 AU in semimajor axis. However, it is not act on the second member of pair (Fig.1). Chaotic zone, related with this resonance is centered at 2.37750 AU, so 5026 Martes can jumped from one side of the exact resonance to other at about 25 kyrs ago.

The main period of the resonant semimajor axis oscillation is about 4000 yr, amplitude is about $5*10^{-5}$ AU. This is the large planet related resonance. It is still present when Yarkovsky effect is taking into account.

It is not present in atlas of Gallardo [10] and some other papers devoted three body resonances [11] (the nearest 3-body resonance is 1 – 4J + 2S at 2.3967AU), so we can conclude that it is not 3-body resonance. The nearest low-order resonance is Jupiter 13:4 resonance (2.3714 AU); it is very distant. Nearest resonance with Mars 20:39M at 2.3782088 AU (delta = 0.00075 AU) have too high order.

To the resonance identification, we have repeated our integration with the only one planet as a perturber and with the very large value of Yarkovsky drift to determinate the real position of resonance. As a result, we obtain that it is single mean motion resonance with Earth, because only in case Earth perturbation we have destruction of the linear evolution of 5026 Martes semimajor axis under the Yarkovsky drift. The position of center of resonance by our numeric integration is about 2.377825 AU (Fig.4).

To study resonance in more detail, we have repeat resonance search for with independent data. We apply Euclid algorithm (Subbotin 1968) **[12]** and obtain that 3:11 resonance with Earth at 2.37757 AU is the most probable main perturbation of 5026 Martes orbit. The result is slightly differ from our integration with only Earth perturbations (2.377825 AU), and from paper Smirnov & Dovgalev (2018) [13] (2.377829 AU).

The next important questions are to explain jumps in semimajor axis of 5026 Martes at about 25 kyr ago. The eccentricity of Martes has maximal value close to 27 kyr (Fig.5). As it is known, the resonance width is proportional to the asteroid and planet eccentricity ratio (Murray, Holman and Potter 1998) [14]:

$$\Delta a \approx 4\alpha \sqrt{\frac{\mu_J e^p e_p^{q-p}}{3\pi(2\varepsilon)^q}} \left(\frac{e_i}{e_p}\right)^{1/2} \sim \left(\frac{e_i}{e_p}\right)^{1/2}$$

This means that the chaotic zone associated with resonance is expanded and can initialize chaotic behavior of the Martes semimajor axis about 27 thousand years ago. Another reason that increases the probability of a chaos is the slow changes in the node of the Earth's orbit and the longitude of the perihelion. Since 27 kyr ago, the precession rate of the Earth's orbit is approximately constant and it is greater than in the previous epoch.

We can outline that resonance is very narrow, not more than with $1*10^{-4}$ AU widths, and has very high order. But we may guess that this resonance is related to the process of forming considered pair.

After the Ceres and Vesta perturbations taking into account, we obtain similar oscillations of semimajor axis with rather smaller period (Fig.6). This means that perturbations of large asteroids slightly reduce the mean semimajor axis of the 5026 Martes but do not change the resonance position. However, it is necessary to take into account close encounter (0.001 AU) with Ceres 3990.34 years ago. Note that the disturbances of Ceres and Vesta do not affect the second asteroid in the pair.

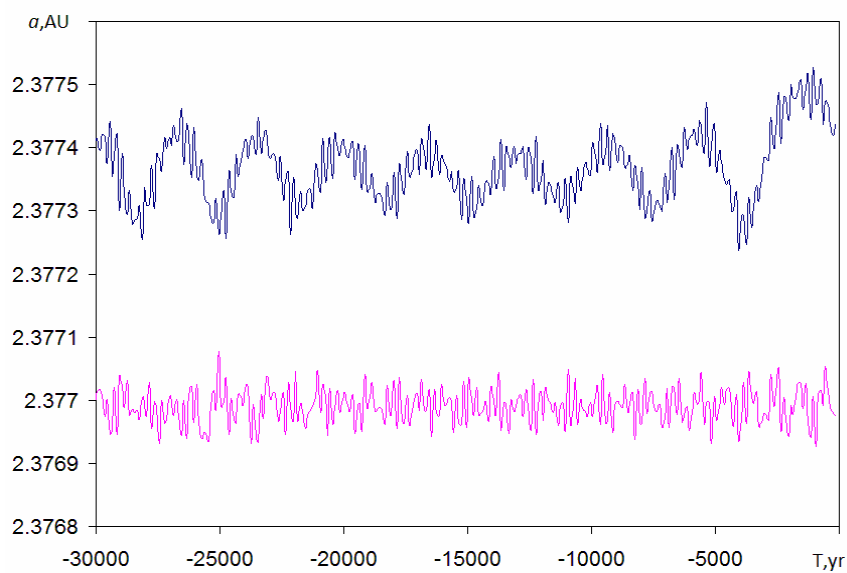

Fig.3. Evolution of semimajor axis **(5026) Martes and 2005 WW113** nominal orbits. All planets perturbations.

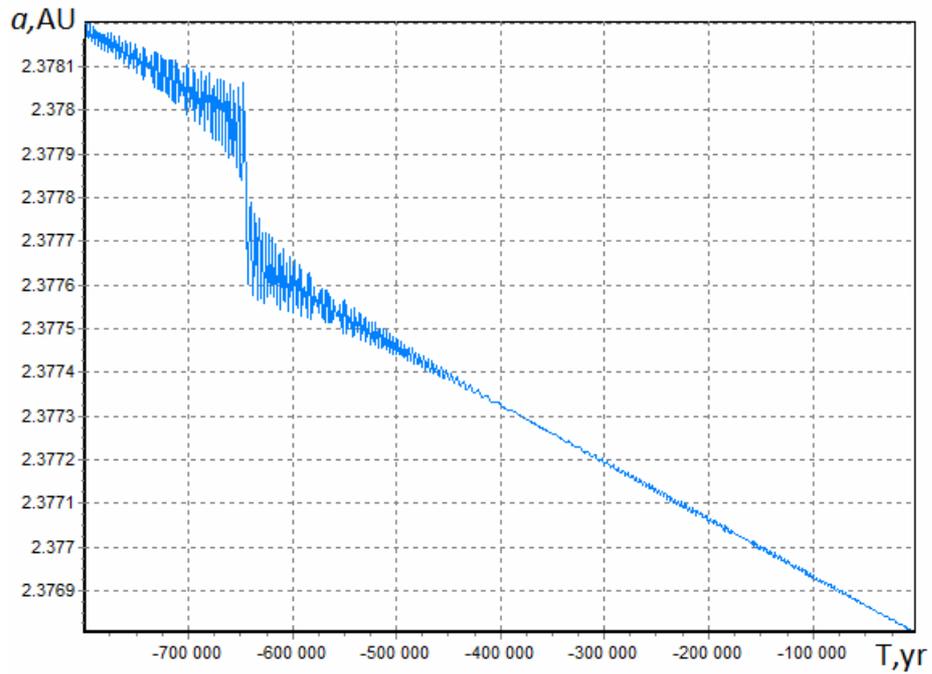

Fig.4. Evolution of the 3026 Martes semimajor axis under only Earth perturbations. Yarkovsky coefficient $A_2=1*10^{-13}$

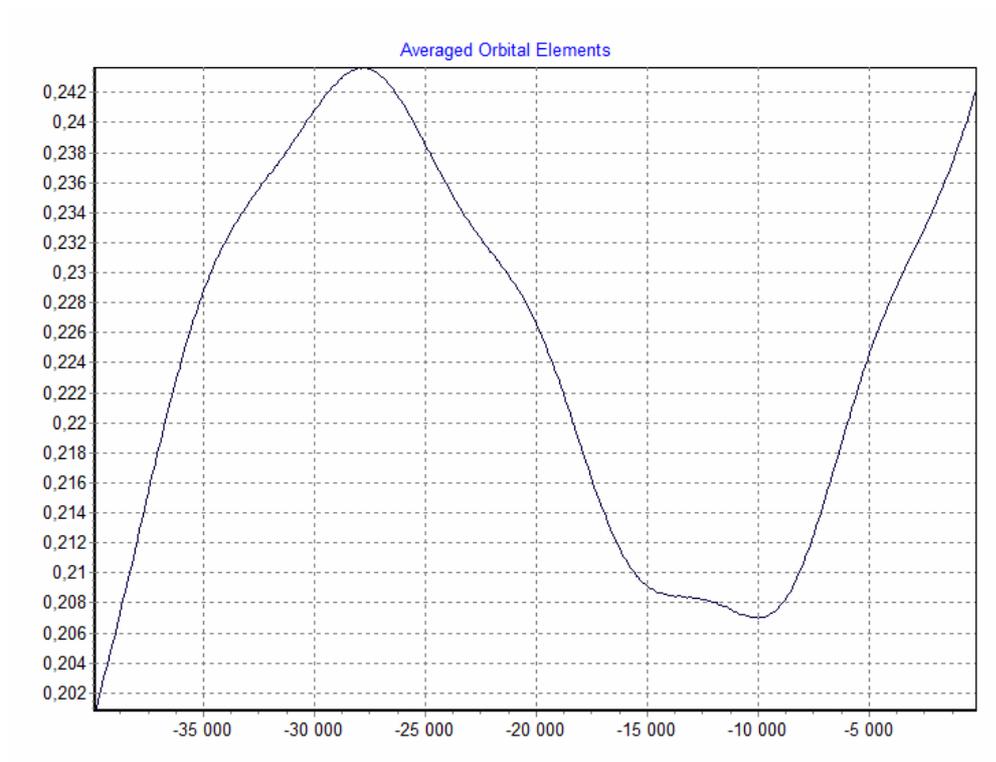

Fig.5. The eccentricity of orbit of 5026 Martes evolution

### The 5026 Martes unstable orbit

The resonant instability of the orbit of 5026 Martes leads to an uncertainty of the orbital position by about 150 degrees (Fig.7) in longitude in the time interval under consideration. As following, the evolution of the semimajor axis for two initial conditions (Table 2 and Table 3) is different (Fig.8). Obviously, this has a big impact on the assessment of the couple's age.

Integrations of the 5026 Martes orbit with different initial conditions show that it is very unstable. Taking into account the perturbations of Ceres and Vesta, we confirm this conclusion. The small variations in the initial orbit or perturbations lead to noticeable changes in the 5026 Martes longitude during the forming encounter. In this context, the careful studying of the resonant perturbations of the 5026 Martes orbit may be important. In principle, it is possible to obtain the dependence of the initial semimajor axis on the average semimajor axis in the fixed considered interval, and then the dependence of the longitude at the end of this interval (i.e. the epoch of the forming encounter) on the initial semimajor axis of Martes.

The long term integration confirms that orbit of Martes is chaotic and very unstable (Fig.9). It confirms a very young age of this pair. The relative velocity rapidly increases since the present epoch up to 1 Myr in the past (Fig.10).

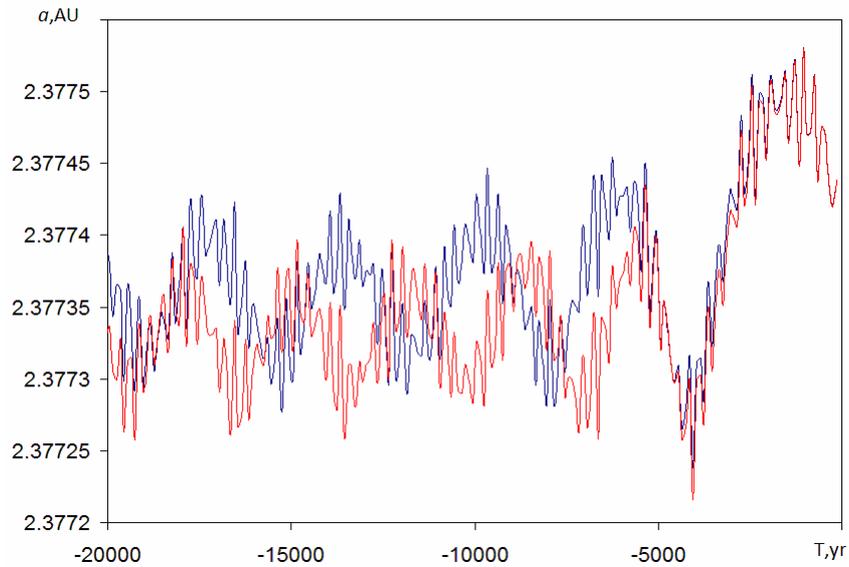

Fig.6. Ceres and Vesta effect (red) and nominal orbit of 5026 Martes (blue)

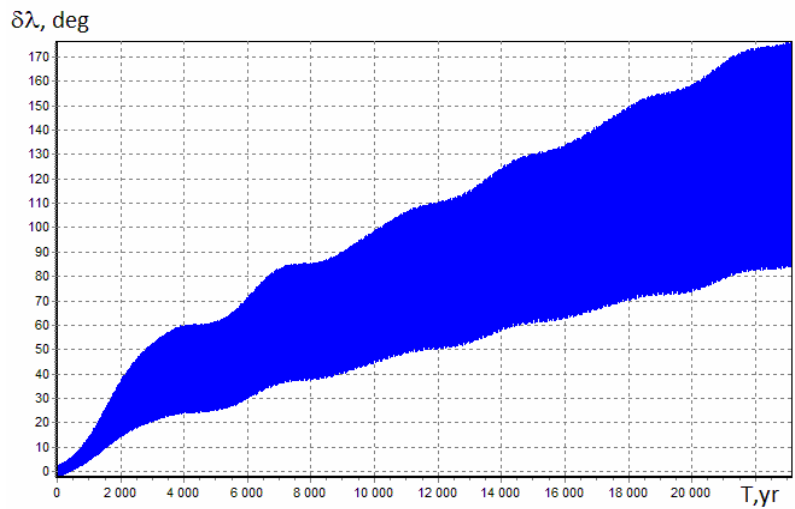

Fig.7. The difference in longitude between two system of initial elements of the 5026 Martes (tables 2 and 3)

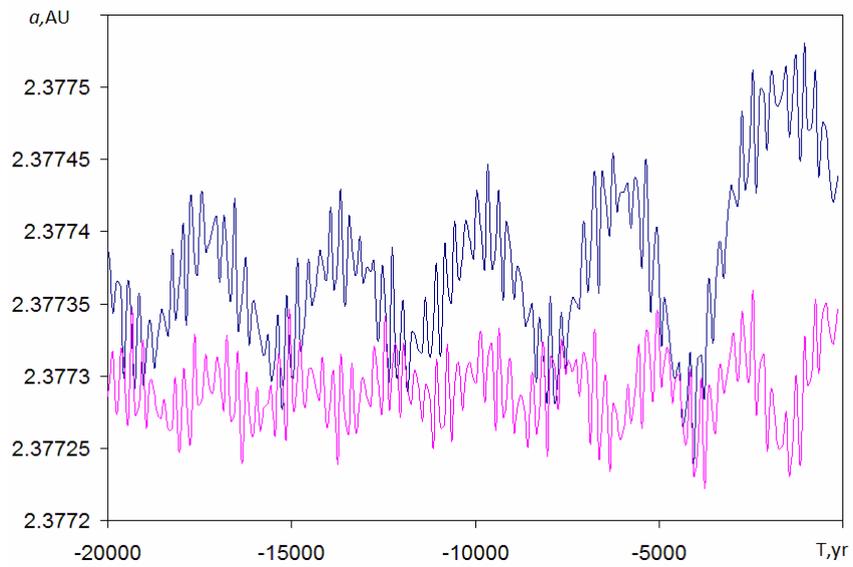

Fig.8. The effect of initial conditions in the semimajor axis of the 5026 Martes
(table 2(red) and table 3 (blue))

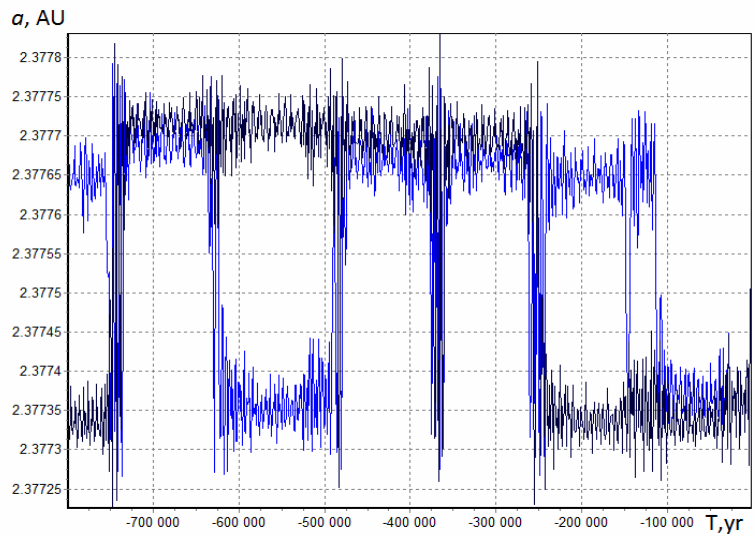

Fig.9. Long time evolution of the 5026 Martes semimajor axis with (blue) and
without (black) Yarkovsky effect

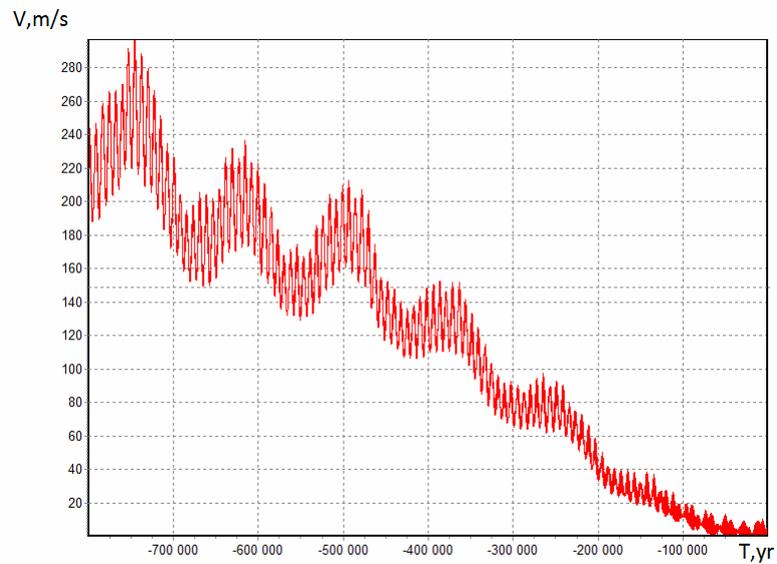

Fig.10. The evolution of the relative velocity in the pair

**Yarkovsky effect**

Currently, the semimajor axis of the (5026) Martes is slightly larger than that of the 2005 WW113 and this difference is of the order of $4*10^{-4}$ AU. At the same time, the diameter of (5026) Martes is significantly larger than the diameter of the 2005 WW113. Accordingly, the expected value of the semimajor axis drift due to the Yarkovsky effect in the 2005 WW113 is greater. Therefore, we can exclude all values of $da/dt_{2005\ WW113}>0$ ($A_2>0$) since the orbits in the pair diverge.

Accept, that the role of accounting for non-gravitational effects in the determination of the age of steam is reduced to a compensation difference of semi-major axes close to the nominal orbit. The expected value of Yarkovsky drift required to eliminate all differences in semimajor axis is **unreal:** $da/dt_{2005\ WW113}=-320*10^{-4}$ AU/Myr.

Our estimation of the Yarkovsky effect parameters is given in the table 5. Our very preliminary studying of Yarkovsky effect shows that distance an encounter can be reduced by more than by order up to 0.0000174 AU. The most probable values $A_2$ for (5026) Martes: $-0.36*10^{-14}< A_2 <0.36*10^{-14}$, for 2005 WW113: $0>A_2>-3.31*10^{-14}$.

Similarly the Ceres and Vesta perturbations, Yarkovsky effect on 5026 Martes change frequency of semimajor axis oscillations. But in addition, it can initialize jumps of 5026 Martes to another side of resonance (Fig.9). After the look on figures above we can to conclude that semimajor axis oscillations and jumps are important attribute of 5026 Martes orbital evolution.

Table 5. Estimated Yarkovsky effect parameters

| Asteroid | H | D, km | r, AU | $A_{21}$, $10^{-14}$ | $\dot{a}\ 10^{-4}$ AU/Myr |
|---|---|---|---|---|---|
| 5026 Martes | 13.9 | 7.52 | 2.377 | **0.30** | **0.9** |
| 2005WW113 | 17.8 | 0.68 | 2.377 | **-3.31** | **-11.1** |

**Conclusions**

The orbital dynamic of very young asteroid pair **(5026) Martes and 2005 WW113** is studied by numerical integration. We discover strong resonant perturbations of the larger member of pair **(5026) Martes** by resonance 3:11 with Earth. This resonance is necessary to account during the study of the pair origin. The orbit of 5026 Martes is much more unstable due to the resonance perturbations. The unbounded secondary (**2005 WW113**) moved far from the resonance and is not perturbed. By this reason, the dependence of the minimal distance and epoch of encounter on the initial orbital elements of 5026 Martes is much stronger. On the other hand, Yarkovsky force has much more effect on 2005WW113 than on 5026 Martes due to size difference.

In general, we have confirmed the results of Pravec et al (2019) [1] about origin of the pair close to 18 kyr. However this epoch has a big uncertainty due to the instability of the (5026) Martes orbit influenced by resonance and due to non-gravity effect on 2005 WW113.